\documentclass[onecolumn,12pt,aps,prd,nobibnotes,showpacs,showkeys,
superscriptaddress]{revtex4-2}
\usepackage{graphicx,graphics,color,epsfig}
\usepackage[colorlinks=true,linkcolor=blue,citecolor=blue,urlcolor=blue,
breaklinks=true]{hyperref}
\usepackage{amsmath,amsfonts,amssymb}
\usepackage{verbatim}
\usepackage{latexsym}
\usepackage{dcolumn}
\usepackage{bm}
\usepackage{natbib}
\usepackage[english,russian]{babel}
\usepackage[utf8]{inputenc}

\begin{document}

\title{\href{}{Estimates of the masses of heavy right-handed neutrino particles using the seesaw mechanism}}
\author{V.~V. Khruschov}
\email{khruschov{\_}vv@nrcki.ru}
\affiliation{National Research Center ``Kurchatov Institute'',\\
Kurchatov~place~1, 123182~Moscow, Russia}
\author{S.~V. Fomichev}
\email{fomichev{\_}sv@nrcki.ru}
\affiliation{National Research Center ``Kurchatov Institute'',\\
Kurchatov~place~1, 123182~Moscow, Russia}

\begin{abstract}
The mass estimates of active and sterile (right-handed) neutrinos are
considered at the phenomenological level using the seesaw mechanism. It is
assumed that the neutrino mass values depend on three characteristic scales,
and the sum of the active neutrino mass values is limited from above by
$0.06$~eV. The neutrino mass values are estimated, as well as the observable
neutrino mass values, namely, the average mass $m_C$ of active neutrinos and
the kinematic neutrino mass $m_{\beta}$ of $\beta$-decay.
\end{abstract}



\maketitle

\section*{INTRODUCTION}

Today, extensive evidence for the existence of dark matter (DM) is available in the gravitational sector of physics. However, the fundamental nature of DM remains unknown (Workman et al., 2022). To explain the nature of DM (see, e.g., Liu et al., 2025), many DM candidates (including those belonging to new physics) have been proposed, such as WIMPs (Weakly Interacting Massive Particles; Feng, 2010; Aprile et al., 2018), ALPs (Axion-like Particles; Marsh, 2016; Irastorza and Redondo, 2018), and PBHs (Primordial Black Holes; Hawking, 1975; Belotsky et al., 2014; Liu et al., 2024). A PBH is a hypothetical object formed in the early Universe, where the energy density is so high that matter could collapse into a black hole. Due to the special formation process, PBHs have an extremely wide range of masses, from the Planck mass of $\sim 10^{16}$~TeV or $\sim 2\times10^{-5}$~g to $\sim 10^5$ $M_{\odot}$ depending on the formation time (Carr et al., 2021; Carr and K\"{u}hnel, 2020). The lower mass limit for DM particles may be around $10^{-22}$~eV (Workman et al., 2022), while the mass of heavy dark matter and superheavy dark matter particles ranges from 1~TeV to the Planck mass (Ibarra and Tran, 2008; Kolb et al., 1999; Pagels and Primack, 1982; Halverson et al., 2017).

Weakly interacting massive particles (WIMPs) are the best candidates for dark matter particles. The masses of WIMPs ($M_{wimp}$) are typically in the range from 10~GeV to 10~TeV (see Feng, 2023). However taking into account the limit for new physics and new particles from paper of Barbieri and Strumia (1999), a slightly larger constraint $M_{wimp}>10$~TeV can be used. It is very likely that heavy right-handed Majorana neutrinos is a component of fermionic WIMPs. They can be related to very small masses of active neutrinos by the seesaw mechanism (see Abdullahi et al., 2023). The seesaw mechanism of generation of left-handed active neutrino masses due to spontaneous symmetry breaking and large masses of right-handed (sterile) neutrinos is discussed in detail for example in the article by Borisov and Isaev (2024) (see also Lindner et al., 2002; Vysotsky, 2011).

\section*{MASSES AND OSCILLATION PARAMETERS OF ACTIVE NEUTRINOS}

Problems of experimental measurement of neutrino masses and theoretical searches for the causes of neutrino masses still remain among the main problems of neutrino physics. Within the Standard Model (SM), neutrinos are considered as massless particles, and this fact in the past corresponded to experimental data. However recently contradictions arose between the data on atmospheric and solar neutrinos, on the one hand, and theoretical calculations, on the other hand. These contradictions were removed after the adoption of the concept of oscillations of massive neutrinos taking into account the Mikheev--Smirnov--Wolfenstein resonances in the processes of neutrino interaction with solar matter.

At present, neutrino oscillations are confirmed by convincing evidence from experiments with atmospheric, solar, reactor and accelerator neutrinos (see Maltoni et al., 2004; Gonzalez-Garcia et al., 2010; Navas et al., 2024; Esteban et al., 2024). Flavor neutrino oscillations can be realized by mixing neutrinos with different masses in the framework of the extended SM with massive neutrinos ($\nu$SM). The squared mass differences $\Delta m_{ij}^2=m_i^2-m_j^2$ and the neutrino mixing parameters are measured in oscillation experiments. However, the absolute values of the neutrino masses cannot be determined in these experiments, as well as the Majorana or Dirac nature of the neutrinos.

Three types of experimental data are sensitive to the absolute neutrino mass scale, namely beta decay data, neutrinoless double beta decay data, and some cosmological and astrophysical data. Each of these data types measures a certain observable neutrino mass. These observable masses are the mean neutrino mass $m_C$, the kinematic neutrino mass $m_{\beta}$, and the effective neutrino mass $m_{2\beta}$, which will be presented below.

The mixing of the three types of light active neutrinos is defined by the Pontecorvo--Maki--Nakagawa--Sakata matrix: $\psi^{\alpha}=U^{\alpha}_i\psi^i$, where $\psi^{\alpha,i}$ are the left chiral fields of flavor or massive neutrinos, $\alpha=\{e,\mu,\tau\}$, $i=\{1,2,3\}$. The Pontecorvo--Maki--Nakagawa--Sakata matrix $U=U_{PMNS}=VP$, the matrix $V$ can be written in the standard parametrization (see Navas et al., 2024) as
\begin{equation}
V = \left(\begin{array}{ccc}
c_{12}c_{13} & s_{12}c_{13} & s_{13}e^{-i\delta}\\
-s_{12}c_{23}-c_{12}s_{23}s_{13}e^{i\delta} &
c_{12}c_{23}-s_{12}s_{23}s_{13}e^{i\delta} & s_{23}c_{13}\\
s_{12}s_{23}-c_{12}c_{23}s_{13}e^{i\delta} &
-c_{12}s_{23}-s_{12}c_{23}s_{13}e^{i\delta} & c_{23}c_{13}
\end{array}\right),
\end{equation}
where $c_{ij}\equiv\cos\theta_{ij}, s_{ij}\equiv\sin\theta_{ij}$,
$P=\{\exp(i\alpha_{CP}),\,\exp(i\beta_{CP}),\,1\}$, $\delta\equiv\delta_{CP}$ is the phase associated with Dirac CP non-conservation in the lepton sector, while $\alpha_{CP}$ and $\beta_{CP}$ are the phases associated with Majorana CP non-conservation.

The currently available parameters of neutrino oscillations (see, e.g., Esteban et al., 2024) at the $1\sigma$ level, determining the flavor oscillations of three light active neutrinos, are as follows:
\begin{subequations}
\begin{equation}
\sin^2{\theta_{12}}= 0.307^{+0.012}_{-0.011}\,,\qquad
\sin^2{\theta_{23}}=\left\{{\rm NO}:\,\, 0.561^{+0.012}_{-0.015}\atop {\rm IO}:\,\, 0.562^{+0.012}_{-0.015} \right.,\tag{2a}
\end{equation}

\begin{equation}
\sin^2{\theta_{13}}=\left\{{\rm NO}:\,\, 0.02195^{+0.00054}_{-0.00058}\atop
{\rm IO}:\,\, 0.02224^{+0.00056}_{-0.00057}\right.,\qquad
\delta/^{\,\circ}=\left\{{\rm NO}:\,\, 177^{+19}_{-20}\atop
{\rm IO}:\,\, 285^{+25}_{-28} \right.,\tag{2b}
\end{equation}

\begin{equation}
\Delta m_{21}^2/10^{-5}\,{\text{эВ}}^2=7.49^{+0.19}_{-0.19}\,,\qquad
\Delta m_{31}^2/10^{-3}\,{\text{эВ}}^2=\left\{\!\!{\rm NO}:\,\, 2.534^{+0.025}_{-0.023}\atop
{\rm IO}:\,\, -2.510^{+0.024}_{-0.025} \right..\tag{2c}
\end{equation}
\label{dat}
\end{subequations}
The values of the $\alpha_{CP}$ and $\beta_{CP}$ phases are currently unavailable from experimental data, as is the absolute neutrino mass scale, for example, the mass value of $m_1$.

Since only the absolute value of $\Delta m_{31}^2$ has been experimentally determined so far, the neutrino mass values can be arranged in two ways:
\begin{equation}
\label{nh}
m_1<m_2<m_3, \qquad m_3<m_1<m_2.
\end{equation}
The first case of the arrangement of neutrino masses in the formula (\ref{nh}) is called as the Normal Ordering of the neutrino mass spectrum (NO), and the second one is called as the Inverted Ordering of the neutrino masses (IO).

It is necessary to experimentally determine at least one value among the observable neutrino masses $m_C$, $m_{\beta}$ or $m_{2\beta}$ to know the absolute scale of active neutrino masses. Here
\begin{equation}
\label{mc}
3m_C=\sum_{i=1,2,3}m_i,
\end{equation}

\begin{equation}
\label{mb}
m_{\beta}^2=\sum_{i=1,2,3}|U_{ei}|^2m_i^2,
\end{equation}

\begin{equation}
\label{mbb}
m_{2\beta}=\bigg|\sum_{i=1,2,3}U_{ei}^2m_i\bigg|.
\end{equation}
The effective neutrino mass $m_{2\beta}$ is the upper diagonal element of the mass matrix $M=U^*m^dU^+$
for Majorana neutrinos, where $m^d={\rm diag}\{m_1,m_2,m_3\}$. The absolute values of the two additional diagonal matrix elements $m_{\mu\mu}$ and $m_{\tau\tau}$ are most likely equal to each other. This assumption does not contradict to some models for the neutrino mass matrix
(see Fritzsch and Xing, 2000; Jora et al., 2010).

The following experimental limits on the observable neutrino masses were obtained at 90\% CL (Esteban et al., 2024): $m_C < 0.013-0.1$~eV (Jiang et al., 2025; Naredo-Tuero et al., 2024; Adame et al., 2025), $m_{\beta} < 0.45$~eV (Acker et al., 2024), $m_{2\beta} < 0.028-0.122$~eV (Agostini et al., 2020; Abe et al., 2024), where the last limit should be increased to $0.079-0.180$~eV to account for the uncertainty in the values of the nuclear matrix elements. Note that the listed constraints on the observable mass values (\ref{mc}), (\ref{mb}) and (\ref{mbb}) do not contradict the previously given values of the neutrino masses and NO of their mass spectrum (Khruschov et al., 2016; Yudin et al., 2016), namely, $m_1\approx 0.0016$~eV, $m_2\approx 0.0088$~eV and $m_3\approx 0.0496$~eV. Moreover, the results of recent cosmological observations (Adame et al., 2025; Naredo-Tuero et al., 2024; Aghanim et al., 2020; Qu et al., 2024; Wang et al., 2024; Sen and Smirnov, 2024) testify in favor of the NO variant of neutrino masses with a total sum of active neutrino masses of about $0.06$~eV.

\section*{SEESAW MODEL AND MASSES OF RIGHT-HANDED (STERILE) NEUTRINOS}

A fundamental question is the nature of the neutrino mass generation. In the absence of a satisfactory theory of this phenomenon, the question can be considered at the phenomeno\-logi\-cal level. First, we assume that there are several different contributions to the neutrino mass, and two of them are most important. We can assume that the first contribution is related to the Majorana mass of the light left-handed neutrino outside the SM. This contribution can arise at some characteristic scale due to the presence in the Lagrangian of an effective term of the Majorana neutrino mass at the modification of the Higgs sector of SM:
\begin{equation}
L'_m=-\frac{1}{2}\overline{\nu}_{L}M_{\nu}\nu^c_{L} + h.c.
\end{equation}
The contribution to the neutrino mass associated with $L'_m$ can be taken into account via the phenomenological parameter $\xi$. The second contribution can be associated with the seesaw mechanism that occurs when heavy sterile (right-handed) neutrinos (HSNs) with masses $M_i$ ($i=1,2,3$) are added to this scheme. This mass contribution can be written as:
\begin{equation}
M''_{\nu}=-M_D^TM_R^{-1}M_D,
\end{equation}
where $M_D$ is the matrix of Dirac neutrino masses and $M_R$ is the mass matrix of sterile neutrinos.

Thus, at least one more new scale appears, which is related to the masses of heavy sterile neutrinos. Let us take a value for $M_D$, which is  proportional to the mass matrix of charged leptons (Borisov and Isaev, 2024; Vysotsky, 2011; Khruschov, 2011):
\begin{equation}
M_D = \sigma m_l,
\end{equation}
where $m_l={\rm diag}\{m_e,\,m_{\mu},\,m_{\tau}\}$, and let us set $M_R\sim M$. Thus, the following phenomenological formula can be used to estimate the neutrino masses:
\begin{equation}
m_{\nu i}=\pm\xi-\frac{m_{li}^2}{M}.
\label{for}
\end{equation}

Using the obtained experimental data (\ref{dat}) on neutrino oscillations, one can find the absolute values of the neutrino masses $\mu_i$ and the characteristic scales $\xi$ and $M$ in eV, for example, in the NO case (Khruschov, 2011):
\begin{equation}
NO: \mu_1\approx0.0693, \; \mu_2\approx0.0698, \; \mu_3\approx0.0851, \;
\xi\approx0.0693, \; M\approx2.0454\times10^{19}.
\label{mnh}
\end{equation}
The estimates of the neutrino masses $\mu_i$ obtained in this way turn out to be too large and do not satisfy the upper limit for the sum of the active neutrino masses, equal to $0.06$~eV. Therefore, we will simplify the mass formula (\ref{for}) to obtain some estimates. If the structure of the Higgs sector
$\nu$SM, consisting only of Higgs doublets, remains unchanged, then the value of $\xi$ can be set equal to zero. Using three values of the masses of heavy sterile neutrinos $M_1$, $M_2$, $M_3$ and some coefficient $\kappa$ (whose value is most likely of the order of unity), instead of the formula (\ref{for}) for the masses of active neutrinos we will use the formula
\begin{equation}
\label{forn}
m_{\nu i}=\kappa{\kern1pt}\frac{m_{li}^2}{2M_i}.
\end{equation}
Substituting the neutrino mass values $m_i$ given above (Yudin et al., 2016; Khruschov et al., 2016) into equation~(\ref{forn}), we obtain the following estimates for $M_1$, $M_2$ and $M_3$ ($1$~EeV = $10^{18}$~eV):
\begin{equation}
M_1\approx 81.6{\kern1pt}\kappa \; {\rm TeV},\quad M_2\approx0.6343{\kern1pt}\kappa \; {\rm EeV},\quad M_3\approx31.8294{\kern1pt}\kappa \; {\rm EeV}.
\label{mne}
\end{equation}
It is likely (if one does not superimpose additional restrictions) that HSNs with masses $M_1$, $M_2$ and $M_3$ can decay in time scales shorter than the lifetime of the Universe, and in this case they are not among components of the stable fermionic dark matter.

One can also obtain  the following estimates of the observable masses $m_C$ and $m_{\beta}$ neutrinos in eV for the case of NO ($m_{2\beta}$ depends on the unknown phases $\alpha_{CP}$ and $\beta_{CP}$):
\begin{equation}
\label{enh}
m_C \approx 0.02, \quad m_{\beta}\approx 0.01,
\end{equation}
which do not conflict with existing restrictions on these mass observables.

\section*{CONCLUSION}

This paper is based on use the known seesaw mechanism of the appearance of small masses of active neutrinos due to large masses of right-handed (sterile) neutrinos (HSNs). A simple formula (\ref{forn}) is proposed for the relationship between the masses of active and sterile neutrinos, which depends on three characteristic scales of HSNs mass values and also some unknown, most likely of the order of unity, coefficient $\kappa$. However, the inverse procedure for finding the masses was used, when the HSNs mass estimates were found based on the phenomenologically estimated values of the active neutrino masses. The presented values of $m_C$ and $m_{\beta}$ (\ref{enh}) may also have practical significance for interpreting future neutrino-related data.

The obtained estimates of the HSNs masses allow us to consider them as candidates for particles of heavy fermionic unstable dark matter. Besides them the class of dark matter particles may include heavy scalar and vector particles that played a role in the formation of the early Universe. Some of them which remain stable compared to the lifetime of the Universe continue to influence the properties of the structural elements of the Universe at the present time as well.

\end{document}